# Size-dependent magnetization switching characteristics and spin wave modes of FePt nanostructures

R. Brandt[1], R. Rückriem[2], D. A. Gilbert[3], F. Ganss[2], T. Senn[4], Kai Liu[3], M. Albrecht[2], and H. Schmidt[1,a]

[1]*School of Engineering, University of California, Santa Cruz, CA 95064, United States.*
[2]*Institute of Physics, Chemnitz University of Technology, D-09107 Chemnitz, Germany.*
[3]*Physics Department, University of California, Davis, CA 95616, United States.*
[4]*Institute of Nanometer Optics and Technology, Helmholtz Center Berlin for Materials and Energy, D-12489 Berlin, Germany.*

**Abstract**

We present a comprehensive investigation of the size-dependent switching characteristics and spin wave modes of FePt nanoelements. Curved nanomagnets ("caps") are compared to flat disks of identical diameter and volume over a size range of 100 to 300nm. Quasi-static magnetization reversal analysis using first-order reversal curves (FORC) shows that spherical caps have lower vortex nucleation and annihilation fields than the flat disks. As the element diameter decreases, the reversal mechanism in the caps crosses over sooner to coherent rotation than in the disks. The magnetization dynamics are studied using optically induced small angle precession and reveal a strong size dependence that differs for the two shapes. Flat disks exhibit well-known center and edge modes at all sizes, but as the diameter of the caps increases from 100 to 300 nm, additional oscillation modes appear in agreement with dynamic micromagnetic simulations. In addition, we show that the three-dimensional curvature of the cap causes a much greater sensitivity to the applied field angle which provides an additional way for controlling the ultrafast response of nanomagnetic elements.

[a]Electronic mail: hschmidt@soe.ucsc.edu

## I. Introduction

High density magnetic nanostructure arrays are of interest for applications such as magnetic random access memory[1-3] and patterned magnetic data storage[4-6]. Individually patterning closely packed elements (e.g. by electron-beam lithography) is time-consuming and expensive. Therefore, the use of self-assembled particles is an attractive alternative. Specifically, the self-assembly of nanospheres produces hexagonally close packed arrays that can be subsequently coated with magnetic material to create a high density array of magnetic elements[7]. The material deposited on the spheres forms curved magnetic "caps" which have an easy axis that is parallel to the surface of the sphere and a radial thickness dependence from center to edge. The curvature of the nanocaps leads to novel features in domain structure and quasi-static switching behavior that have been studied extensively[8-13]. For example, Albrecht *et al*[14] deposited ferromagnetic, single-domain Co/Pd multilayers onto self-assembled nanospheres, resulting in quasi-static switching behavior that deviated from simple Stoner-Wohlfarth switching as a result of the relative geometry of the applied field angle and the varying easy axis direction and strength across the cap. In another study, Makarov *et al*[15] found that the interconnecting material between the spheres did not magnetically couple the caps, but served as domain wall pinning sites, creating a percolated medium. So far, most studies of caps have focused on static magnetic properties, but recently it was shown that the spin wave modes in 100nm diameter FePt caps show distinct differences compared to flat nanomagnets of equal dimensions[16].

Here, we report on a comprehensive study of the size-dependent magnetic properties of both curved and flat FePt nanomagnets. The quasi-static switching behavior is analyzed using the first-order reversal curve (FORC) method. We find that large magnets of both shapes switch by vortex annihilation, but the spherical caps switch to a coherent rotation mechanism at a larger diameter than the flat disks. To measure the intrinsic spin wave modes, single element measurements were made using a time-resolved optical pump-probe experiment on three different nanomagnet diameters. Not only do the spherical caps show many more oscillation modes compared to the flat disks for all sizes, but also the frequency spectrum becomes more complex as the diameter increases. Both quasi-static and dynamic measurements are validated by micromagnetic simulations. Furthermore, we show that these demagnetizing regions in the cap can be manipulated by changing the applied field angle relative to the substrate plane, leading to oscillation mode spectra that are far more sensitive to the field angle than in a flat disk.

**II. Sample Preparation**

A series of close packed arrays of $SiO_2$ nanospheres with three different diameters (100, 160, and 300 nm) was created by the method proposed by Micheletto *et al*[17] on thermally oxidized silicon substrates (Fig. 1(a)). This approach involves a self-assembly process which is mainly driven by capillary forces in an aqueous colloid, where the water is evaporating under controlled conditions (temperature and humidity). A 20 nm thick chemically disordered $Fe_{55}Pt_{45}$ film and 2 nm thick Pt capping layer were deposited at room temperature on the nanosphere array by DC magnetron co-sputtering from pure element targets at an Ar pressure of $3.5\times10^{-3}$ mbar. The iron content of (55±1) at. % in the alloy film was determined by Rutherford backscattering spectroscopy (RBS). The total magnetic moment of an unpatterned film sample (~10 $mm^2$) was determined by superconducting quantum interference device - vibrating sample magnetometry (SQUID-VSM), and was divided by the volume of the alloy to calculate a saturation magnetization of $\mu_0 M_S = 1.27$ T, which is in good agreement with published values[18]. The film ideally has a uniform thickness in the direction perpendicular to the substrate, but the curvature of the sphere's surface causes a radial change in the actual thickness from center to edge (Fig. 1(a) cartoon cross-section). The material which deposits between the spheres constitutes only 9% of the sputtered material and is considered as a minor perturbation to the measured magnetic signal (and even does not result in a measurable signal for magneto-optic measurements).

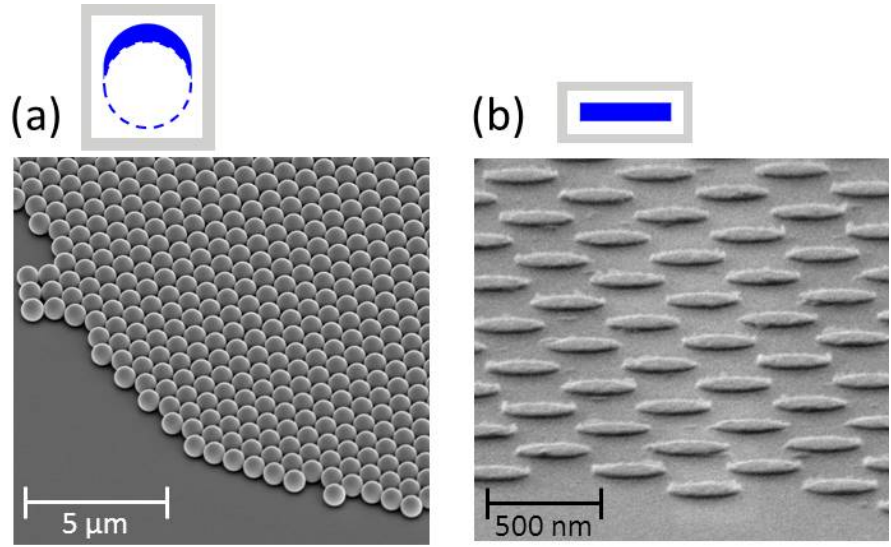

Fig. 1. (Color online) SEM images of (a) a self-assembled monolayer of spheres and (b) a patterned array of flat, circular disks with a 300 nm diameter (images taken at oblique angle). The cartoon cross-sections above show the geometry of the magnetic elements.

A series of flat magnetic disks with identical composition, total volume, out-of-plane thickness, and in-plane diameter to the spherical caps was fabricated by standard electron beam lithography and lift-off on a planar silicon substrate with a nonmagnetic antireflective (AR) coating (Fig. 1(b)). Both isolated magnetic elements as well as nanomagnet arrays with a center-to-center distance of twice the disk diameter were fabricated. The AR-coated substrate (designed for a wavelength of 800 nm) improves the signal-to-noise ratio of magneto-optical measurements on single nanostructures by reducing probe beam reflections from the area surrounding the investigated disks[19].

For quasi-static switching measurements, arrays of nanoelements (discussed in the next section) were measured while dynamic measurements were taken on single nanostructures. To isolate a spherical cap, sample areas that were sparsely populated with spheres were identified. Then, focused ion-beam milling was used to remove the surrounding magnetic film to an outer diameter of 6-8 μm, which is larger than our probe beam spot size (see Fig. 5(b) for an SEM image). Being able to probe individual magnetic nanostructures is essential to measuring the intrinsic magnetic properties.

**III. Quasi-static reversal behavior**

**A. First-order reversal curve method**

The first-order reversal curve (FORC) method is a technique used to study the switching behaviorr of hysteretic systems by probing the minor hysteresis loop spectrum instead of just the major loop. This analysis can be used to qualitatively and quantitatively "fingerprint" magnetization reversal[20-22]. The FORC method is performed as follows: First, the sample is saturated in a positive field. Then, the field is reduced to a defined reversal field, $H_R$, and the magnetization is recorded as the applied field, $H$, is swept back to positive saturation, hence tracing out a single FORC. This sequence is repeated for decreasing values of the reversal field until negative saturation is reached, measuring a family of FORCs where the magnetization, $M$, is recorded as a function of both $H$ and $H_R$. The FORC distribution, $\rho$, is then calculated by applying a mixed second order derivative:

$$\rho(H, H_R) = -\frac{1}{2}\frac{\partial^2 M(H, H_R)/M_S}{\partial H \partial H_R} \qquad (1)$$

where $M_S$ is the saturation magnetization. The resulting distribution is only nonzero for irreversible switching processes[23].

To measure the FORCs on the spherical caps, sample pieces (~ 3 × 3 mm$^2$) were cut and measured by a vibrating sample magnetometer (Princeton Measurement Corp. 2900 VSM). For the 20 × 20 μm$^2$ arrays of flat disks, FORCs were taken using a longitudinal magneto-optical Kerr effect magnetometer (Durham NanoMOKE-2) with a 632.8 nm HeNe laser and a spot size of ~20 μm. For both sets of measurements, the field was applied in the plane of the substrate, and FORC loops were recorded with equal $H_R$ spacing (~1 mT) for a field range of +25 mT to -70 mT.

## B. Experimental Results

Fig. 2 shows the families of FORCs (insets) and the corresponding distributions for both the flat disks (Figs. 2(a)-(c)) and spherical caps (Figs. 2(d)-(f)) for all three nanomagnet diameters - 100, 160 and 300 nm. The outer boundaries of the families of FORCs delineate the major loops (Fig. 2 insets). For all the flat disks, the major loops are highly pinched, with little remanence, which is characteristic of magnetization reversal via a vortex state[24]. In the smallest 100 nm disks, the nucleation field, signified by the abrupt drop in magnetization towards the flux closure state, is significantly smaller than in the larger disks; the remanence is appreciably larger. The larger spherical caps with 160 and 300 nm diameters also exhibit highly pinched loops (insets in Figs. 2(e) and 2(f)). The magnitudes of the vortex nucleation and annihilation fields are much smaller than their flat disks counterparts. Note that for all the pinched loops, the FORCs with $H_R$ near zero essentially overlap and display a linear magnetic field dependence, corresponding to the reversible motion of the vortex cores inside the nanomagnets. Interestingly, the loop from the 100 nm diameter spherical caps has a different shape from the others, with no appreciable loop pinch and a much larger remanence of ~ 50%. The FORCs fill the interior of the major loop rather uniformly, in sharp contrast to the other samples. These characteristics suggest the presence of a different reversal mechanism.

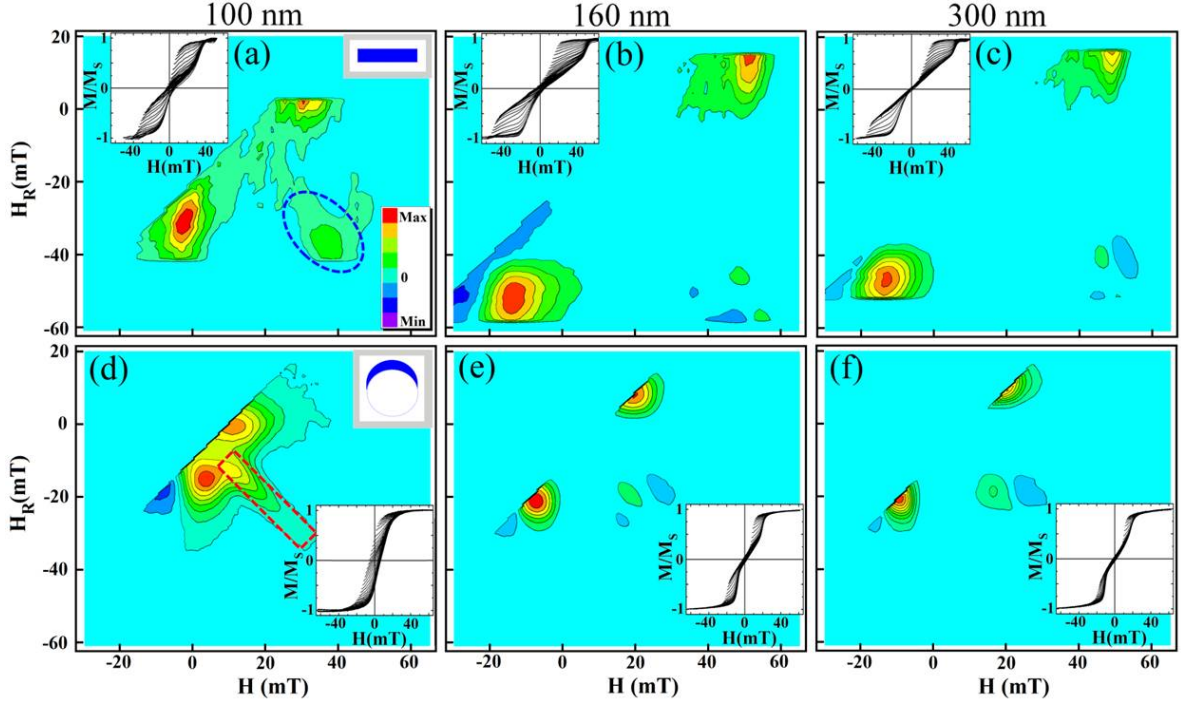

Fig. 2 (Color online). FORC distributions for (a-c) flat disks and (d-f) spherical caps. The FORC curves are shown in the inset, and the diameter is listed above each column. The dashed circle in (a) highlights a FORC feature caused by shape asymmetry, while the dashed box in (d) represents a FORC distribution from coherently reversing caps.

Turning to the FORC distributions, all the flat disk distributions show predominantly a two lobe feature in the upper right and lower left corners of Figs. 2(a)-(c). They correspond to the irreversible nucleation and annihilation events of the vortex reversal, as shown in earlier studies[20,25]. There is also a shallow third feature located in the lower right of each panel, due to the small amount of shape asymmetry in the disks[20]. This feature, indicated by a dashed circle in Fig. 2(a), is more pronounced in the 100 nm disks than in the larger disks, indicating a larger impact of shape asymmetry in the smaller disks. Similarly, the 160 and 300 nm diameter caps also predominantly show a two lobe structure, occurring at much smaller values of $H$ and $H_R$. The decreased values mean that the vortex nucleation and annihilation fields are much smaller in the caps as compared to the disks. This indicates that the energy benefit associated with the vortex state decreases more rapidly with field in the caps than in the disks, and suggests that the demagnetization energy in the caps plays a stronger role and evolves differently compared to the flat disks.

Interestingly, the FORC distribution for the 100 nm diameter caps also has a two-lobe feature centered at ($H$ = 14 mT, $H_R$ = 2 mT) and ($H$ = -3 mT, $H_R$ = -15 mT), in addition to a ridge along the $H$ = -$H_R$ axis, which is much more pronounced and occurring at lower values of ($H$, $H_R$) than the aforementioned feature caused by shape asymmetry. This ridge, indicated in Fig. 2(d) by a dashed box, has been previously shown to correspond to reversal by coherent rotation[20], indicating that the 100 nm spherical caps reverse *both* by vortex nucleation/annihilation and by coherent rotation. By integrating the FORC distribution over the ridge feature, we estimate the fraction of disks reversing by coherent rotation[26,27] to be 17%, while the remainder reverses by vortex nucleation/annihilation.

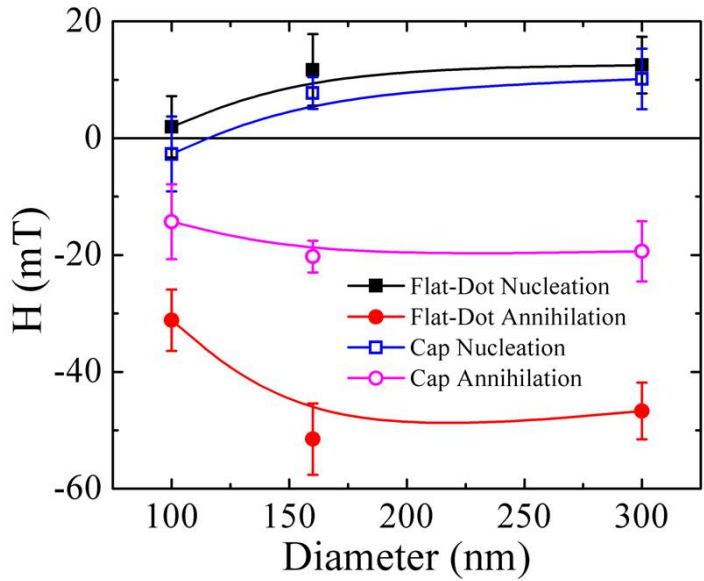

Fig. 3 (Color Online). Nucleation (solid) and annihilation (open) fields for both the flat disks (squares) and spherical caps (circles) as extracted from the FORC distribution along a decreasing-field sweep. Lines are included as guides and are developed from a B-spline fit.

Values of the vortex nucleation field ($H_N$) and annihilation field ($H_A$) can be extracted from the central positions of the lobe features, as described earlier[20,25]. Consider a vortex reversal process from positive to negative saturation, with $H_N$ being the nucleation field coming from positive saturation, and $H_A$ corresponding to the negative annihilation field (achieving negative saturation). All $H_N$ and $H_A$ values are plotted in Fig. 3. In both cases, the nucleation and annihilation fields are much smaller for the 100 nm disks/caps compared to the 160 or 300 nm diameters. The increase in $H_N$ and $H_A$ with increased diameter confirms the increased stability of the vortex state in the larger disks. Comparing the disks with the caps, we quantitatively confirm what is suggested by the *M-H* loops, specifically that the curved caps have significantly reduced nucleation and annihilation fields. This demonstrates that the curved structures indeed facilitate magnetization reversal via a vortex state.

As shown in Fig. 3, the annihilation field is affected more strongly by the curvature than the nucleation field. Specifically, the annihilation field in the caps is reduced by an average of 25 mT, while the nucleation field is reduced by an average of only 4 mT. This suggests that the energy barrier to nucleate a vortex is similar between the caps and disks, however, in the caps, the demagnetization energy increases much more rapidly as the field is increased towards $H_A$. For both the 100 nm disks and caps, a non-zero remanence exists as a result of the small (and even negative) $H_N$ values and associated nucleation field distributions. The caps in the single domain state further increase the overall remanence to ~ 50% as shown in Fig. 2(d) inset. Thus, for the spherical caps, the 100 nm diameter is at the boundary of the two reversal modes, while this crossover is suppressed to below 100 nm in the disks. Thus, the FORC analysis enabled us to identify the mixed reversal modes.

**C. Simulations**

To interpret these results, and understand how the curvature induces coherent reversal at a larger diameter, micromagnetic simulations were performed using the Magpar Micromagnetics package[28]. The numerical simulations are based upon the Landau-Lifshitz-Gilbert (LLG) equation, which describes the motion of the magnetization:

$$\frac{d\vec{M}}{dt} = -\gamma_0\mu_0\vec{M} \times \vec{H}_{\text{eff}} + \frac{\alpha}{M_S} \tag{2}$$

where $\gamma_0$ is the electron's gyromagnetic ratio, $\alpha$ is the Gilbert damping constant, $\mu_0$ is the permeability of free space, and $H_{eff}$ is the effective field that represents various energy contributions in the system, such as shape or crystal anisotropy. We used the measured saturation magnetization ($\mu_0 M_S$ = 1.27 T) and an exchange constant of $10^{-11}$ J/m, and assumed that there is no crystalline anisotropy – corresponding to chemically disordered FePt. To simulate a loop from positive to negative saturation, the magnetization is first relaxed at the highest field strength with the Gilbert damping parameter set equal to one. This magnetization state is then the starting point for the next applied field strength. The magnetization reaches equilibrium at each step over the field range, and the x-component (defined in Fig. 4(a)) of the magnetization at equilibrium is plotted.

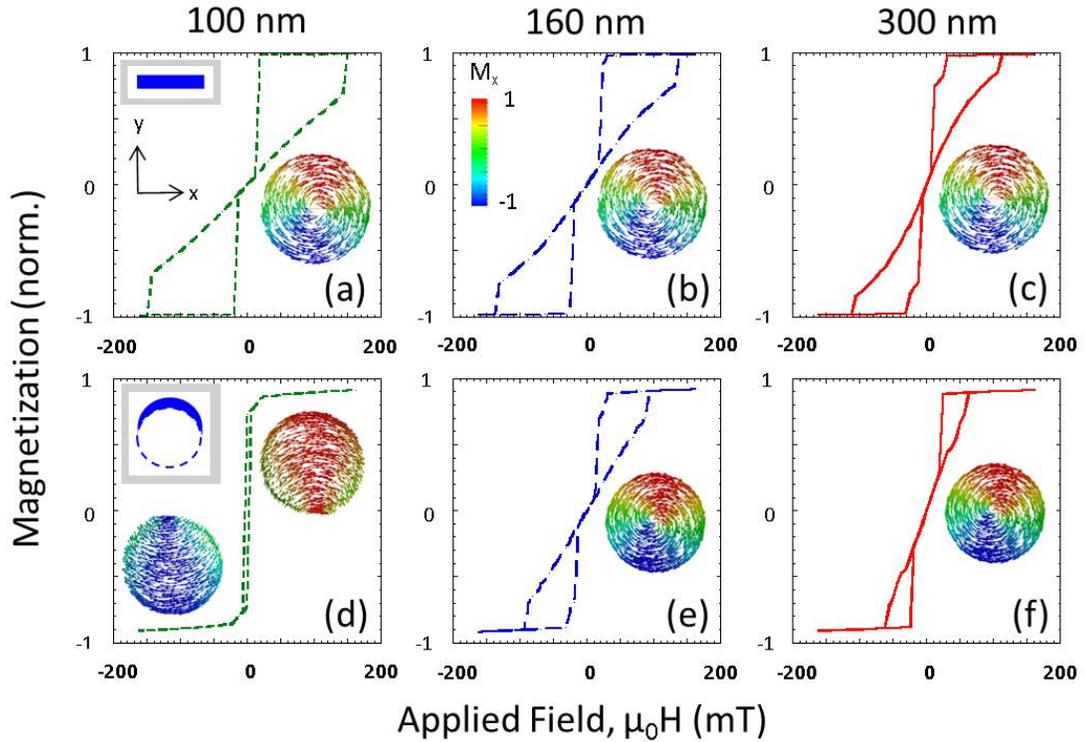

Fig. 4. (Color online) Simulated major hysteresis loops for single (a-c) flat disks and (d-f) spherical caps. The sizes on the top correspond to the diameters of the elements, and the insets show the magnetization equilibrium at zero applied field (the colorbar in (b) corresponds to the x-component, defined in (a)). The insets in (d) show the equilibrium magnetization just before and after the coherent reversal. The simulations were carried out for a temperature of T=0 K.

The simulation results for single magnetic elements are shown in Fig. 4, and the switching mechanism for each case matches the experiment. It should be noted that the x-axis scale for the applied field is three times that of the experiment (65 mT vs. 200 mT). The simulations assume a temperature of zero degrees Kelvin, therefore, the applied field needs to be higher for switching since there are no thermal fluctuations or defect sites which aid in the process. Nevertheless, the qualitative comparison of the hysteresis loop shape and relative saturation fields is valid. For the disks, switching occurs via vortex nucleation and annihilation, while for the caps, only the two larger sizes switch via vortex and the 100 nm caps exhibit coherent rotation reversal. As we learn from the FORC distribution, the 100 nm caps exhibit both types of switching, and this is due to the size distribution of the sphere samples. For the 100 nm diameter sample, the sphere sizes range from 90 – 130 nm diameter (as determined by SEM analysis), and simulations show that for diameters > 110 nm, vortex switching occurs. In addition, the simulation correctly predicts that the vortex nucleation/annihilation fields are lower in the spherical caps. Since we are simulating single elements (and ignoring dipolar effects of the array), the smaller annihilation fields in the caps can be attributed directly to the three-dimensional shape.

One of the significant features of the spherical caps is that the radial thickness tapers from center to edge. This change in thickness can explain both the change to coherent rotation at a larger diameter and the reduced annihilation fields in the 160 nm and 300 nm caps. For a constant thickness, the critical radius between the switching mechanisms (vortex nucleation/annihilation vs. coherent rotation) is related to the balance between the exchange energy, which favors alignment of the spins, and the shape-dependent demagnetizing energy, which favors the reduction of poles on the surfaces[29]. In the same way that there is a critical radius, there is a critical thickness that separates the two switching mechanisms[30,31]. The radial thickness gradient across the cap gives an overall effective thickness that is less than the flat disk thickness, and thus, the caps cross over to coherent rotation at a larger diameter.

In the same way that the tapered radial thickness of the caps causes a crossover to coherent rotation in the nominally 100 nm caps, the thinness of the spherical cap edges is also responsible for the suppression of the annihilation field in the 160 nm and 300 nm diameter caps. As the reversal field is applied, the vortex travels towards the edge. The reduced thickness at the edge makes the vortex unstable (again related to the critical thickness), and therefore, it is energetically favorable to align the spins and saturate. Thus, we observe that the curvature of the caps changes the switching behavior considerably as compared to disks of the same diameter. The curvature also results in a larger effective radius – as measured along the surface of the sphere – which tends to favor reversal by vortex nucleation/annihilation. However, in this case, we find the thickness gradient to play a more dominant role in the reversal mechanism. In the next section, we will see that the curvature affects the magnetization dynamics drastically as well.

## IV. Spin wave modes

### A. Time-resolved magneto-optical Kerr effect magnetometry

In order to measure the ultrafast magnetization dynamics, we use an all-optical time-resolved magneto-optical Kerr effect (TR-MOKE) microscope[32,33]. The ultrafast pulses from a Coherent MIRA Ti:Sapphire laser are split into pump (frequency-doubled 400 nm, ~10.6 mW/μm$^2$) and probe (800 nm, ~2.65 mW/μm$^2$) beams. The heating of the relatively strong pump beam causes an ultrafast demagnetization followed by a small-angle precession around the applied field[34]. The probe beam travels through an optical delay line and linear polarizer before being focused to a spot size of ~1 μm onto the pumped area using a 60× microscope objective (N.A. = 0.85). The change in polarization of the probe beam is measured using a crossed-polarizer configuration, and is recorded as a function of delay time between the pump and probe pulses. Another path analyzes the change in reflectivity (no polarizer before the detector), and is used to identify nonmagnetic signals present in the measurement. The applied field is provided by small permanent magnets, which produce magnetic fields with both in-plane and out-of-plane components. The resultant angle of the applied field is (30 ± 2)° from the substrate plane, and its strength is varied from $\mu_0 H$ = 0.25 T to 0.55 T. It is important to note that the nanomagnets are saturated for all applied field strengths, independent of shape, as verified by the hysteresis loop measurements in the previous section.

To extract the magnetic response of the nanomagnets from the time trace data, the thermally induced background is eliminated by subtracting a monoexponential decay from the raw signal in both the crossed-polarizer (magnetic) and reflectivity (non-magnetic) channels[35]. Then, a fast Fourier transform (FFT) is taken, and the spin wave frequencies that have magnetic origin are identified by comparing the spectra for the two channels. We fit a sum of Gaussians to the identified magnetic peaks (determined over several scans), and the error in fitting the peak position gives the value of the error bars for the experimental data.

**B. Experimental results**

Fig. 5 shows example data taken for one field strength ($\mu_0 H$ = 0.41 T) for all diameters of flat disks and spherical caps. The left columns show the background subtracted time traces, and the right columns are the corresponding Fourier spectra (Gaussian peak fits shown for identified magnetic frequencies for the smallest elements, $d$ = 100 nm). The disk spectra (Fig. 5(a)) exhibit two identified magnetic modes, but the spherical cap spectra (Fig. 5(b)) are considerably different, showing more than two spin wave modes for every size. The additional peaks in the disk spectra (e.g. the significant peak near 5 GHz for $d$ = 100 nm) are not magnetic signals since they also appear in the reflective channel, and hence, represent noise in the measurement.

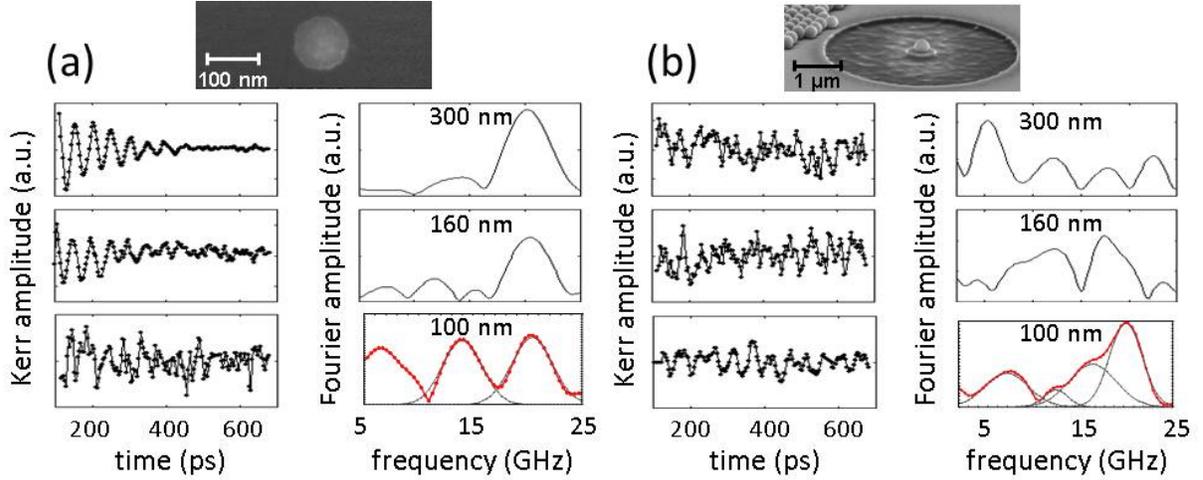

Fig. 5. Experimental data taken at $\mu_0 H = 0.41$ T for (a) flat disks and (b) spherical caps at three diameters. The left columns are the background subtracted time traces, and the right columns are the corresponding Fourier spectra (Gaussian fits included for the $d = 100$ nm data). The SEM images show (a) an individual 100 nm disk and (b) a single 300 nm sphere that has been isolated using ion beam milling.

Fig. 6 displays the extracted peak oscillation frequencies (symbols with error bars) as a function of applied field for all three sizes of flat disks (a-c) and spherical caps (d-f), where the diameter is listed above each column. For the disks, two main oscillation modes are detected for each size: The higher frequency mode changes only slightly as a function of diameter, but the lower frequency mode noticeably decreases as the diameter is increased. Clear edge modes were not recorded for the lowest and highest magnetic fields for the 300 nm diameter disk. While the disks only exhibit two magnetic oscillation modes, the frequency spectra of the caps are considerably more complex with 3-4 observed modes for each size. All the caps contain at least two modes whose frequencies increase with increasing applied field, and at least one mode that does not vary with applied field, in stark contrast with the flat disks.

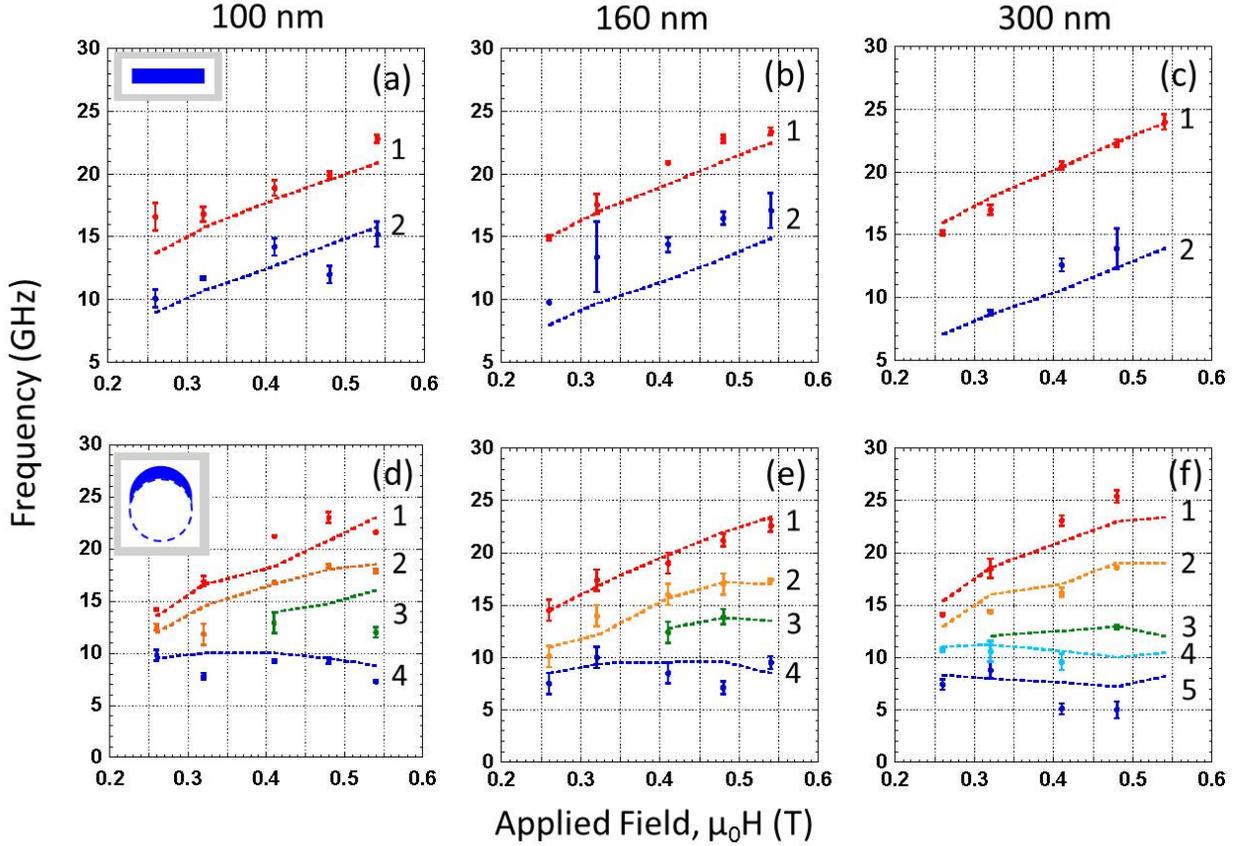

Fig. 6. (Color online) Experimental data (symbols with error bars) and micromagnetic simulation (dashed lines) for three diameters of (a-c) flat disks and (d-f) spherical caps. The diameter is listed above each column. The external field is applied at $(30 \pm 2)°$ from the substrate plane.

**C. Micromagnetic simulations**

The experimental data taken on the disks and caps show very different frequency spectra. This qualitative behavior has previously been observed and discussed for the smallest elements of 100 nm diameter[16]. It was found that the two modes in the disks correspond to the well-known center and edge modes[36] that arise due to the inhomogeneous demagnetizing field ($H_d$) within a confined element, while the cap breaks up into multiple regions with distinct demagnetization fields whose magnitude depends on the relative magnetization orientation of the cap section with respect to the external field[16].

Here, we focus on the size dependence of this behavior. To this end, dynamic micromagnetic simulations based on the LLG equation were run using the FEMME[37] package, assuming no crystalline anisotropy, an exchange constant of $10^{-11}$ J/m, and using the measured $M_S$ value. To explore oscillations in the linear regime, the samples were first brought to a relaxed saturated state, in a constant external field (matching the experimental conditions), then a fast (100 ps) in-plane perpendicular pulse field of 1 mT was applied. Magnetization dynamics were then simulated over the next 4.0 ns; a damping factor of $\alpha = 0.01$ was used. The resulting oscillation is analyzed for its peak frequencies via FFT. To visualize the spatial mode distribution, the sample was again brought to a relaxed saturated state in a constant external field, and then one spin wave was excited with its precession frequency by an alternating field, which was applied perpendicular to the constant external field in the in-plane direction[38]. This step produces the mode image, which was extracted by comparing minima and maxima of the z-component of the magnetization value of every finite-element mesh-node during a simulation time of 10.0 ns (chosen to ensure at least 10 oscillations).

The simulated peak frequencies are plotted as dashed lines in Fig. 6, and the agreement with the experimental data for this large range of sizes and fields and using a single set of (un-fitted) parameters is rather good. For the flat disks, the two oscillation modes correspond to the previously observed center and edge modes for all diameters (see mode image in Fig. 7(b)). We find that, as the disk diameter is increased, the frequency of the center mode increases while the frequency of the edge mode decreases, in agreement with the experimental results. Similar observations have been made in non-ellipsoidal nanomagnets[36,39]. The center mode frequency increases with diameter because the magnitude of the demagnetizing field decreases ($H_d$ lowers $H_{eff}$) as the disk becomes less confined and approaches a thin film. The edge mode, on the other hand, decreases in frequency due to the reduced curvature at the edge of the larger disks. The demagnetizing field in the spin wave wells is sensitive to the shape of the edge region, thus as the radius of curvature changes, so do $H_d$ and the resonance frequency. This interpretation is consistent with a study on asymmetric egg shapes[40], in which the varying radius of curvature on the two ends of the egg caused a splitting of the edge mode as a direct result of the different shape anisotropy at the two edges. The same effect is at work in our study where larger disk diameters correspond to smaller curvatures and thus lower resonance frequencies. The fact that the 160 nm edge mode has a very consistent high-frequency offset from the simulation at all applied fields indicates an intrinsic imperfection which changed the internal field, most likely fabrication defects near the disk edge. This is consistent both with the SEM images of the disks and with the fact that the center mode, which is insensitive to edge imperfections, matches the simulations very well.

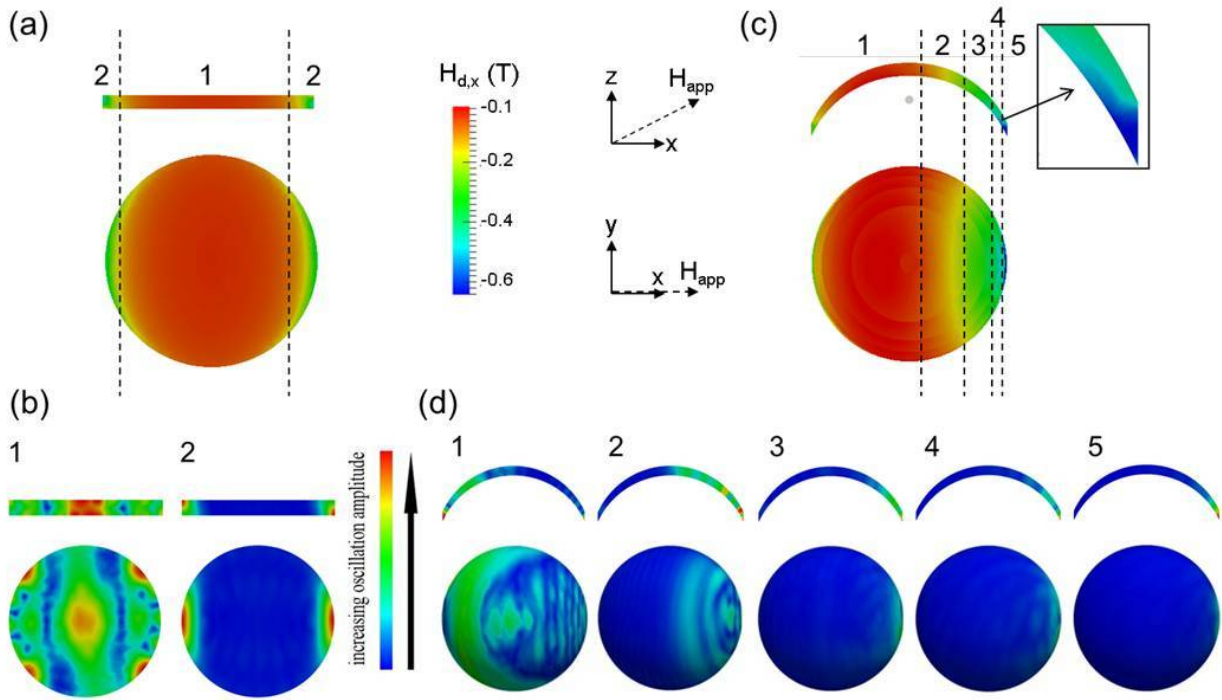

Fig. 7. (Color online) Simulated demagnetizing field (a/c) and mode images (b/d) for a 300 nm diameter (a/b) flat disk and (c/d) spherical cap ($\mu_0 H_{app}$ = 0.54 T). The x-component of the simulated $H_d$ is shown in (a) and (c) for both side and top-down views with the applied field direction and color scale indicated in the center. The mode images in (b) and (d) (numbers correspond to the frequency branches in Fig. 6) are normalized from blue (minimum oscillation amplitude) to red (maximum).

In the same way that the shape anisotropy, or demagnetizing field, creates the localized center and edge modes in the disks, we expect that the three-dimensional curvature of the spherical caps is responsible for their more complex dynamic behavior. The key difference between the cap and disk is that the demagnetizing field distribution of the cap is asymmetric. This arises due to the local film plane orientation across the cap and the tilted applied field direction, which produces different demagnetizing fields in those regions. In order to see if this behavior is modified by the change in diameter, Fig. 7(c) shows the simulated x-component of $H_d$ for the largest ($d$ = 300 nm) diameter cap. Both the top-down and side views are shown (with the color bar and relative applied field orientation in the center). On the left side, where the local film plane is aligned with the applied field, $H_d$ is much smaller than on the right side, where the applied field is pushing the magnetization out of its easy axis direction. Just as $H_d$ is asymmetric, the mode images are correspondingly asymmetric (Fig. 7(d)). Furthermore, since the curvature of the cap and the applied field direction create several distinct regions of demagnetizing magnitude, there are an increased number of oscillation modes as compared to the disks. In the case of the 300 nm diameter caps, since the magnetic element is larger than previously studied ones, it can sustain a larger number of modes, up to five under the applied conditions. Modes 1 and 2 are generally localized to the left and right sides, respectively, while modes 3-5 are highly confined to the far-right side and correspond to the varying regions of $H_d$: mode 3 being confined to green, mode 4 to light blue, and mode 5 to dark blue (close up of edge inset in Fig. 7(c) to show light blue region). These highly localized modes correspond to the "pinned" low frequency modes seen in Fig. 6(f) (modes 3-5) that show very little field dependence. They are quite close in frequency, and in our experiment, we only observe two of the pinned modes at any given time, most likely due to the limited time scan of our apparatus and thereby limited frequency resolution. The atypical field dependence of these low frequency modes is due to the tapering of the thickness at the edges which causes the magnetization to more strongly prefer the local in-plane orientation. Near the right edge, the applied field is nearly perpendicular to the local film plane, and therefore the demagnetizing field is maximized in these areas. If $H_d$ is extracted (not shown), it is seen that the magnitude of $H_d$ increases as $H$ increases, and they effectively cancel each other, causing a flat field dependence. Although we have shown the mode images for the 300 nm diameter case, the 100 nm and 160 nm give similar asymmetric responses. However, the increase in the number of "pinned" modes as the diameter is increased is due to the larger size of the cap which allows additional localized oscillation regions.

As evident from the previous analysis, the interplay between the applied field direction and the local film plane along the spherical cap produces a more complex frequency spectrum. This raises the question of how the magnets would respond if the applied field angle were varied. For example, if the field were applied in the plane of the substrate – as was the case for the quasi-static measurement – then the left and right sides of the cap would be symmetric with respect to the field geometry. Fig. 8 shows the simulated demagnetizing field distribution (left) and simulated frequency spectrum (right) for a 100 nm diameter cap at an applied field strength of $\mu_0 H$ = 0.54 T at three different field angles. The total simulation time (1.5 ns) exceeded the experimental scan time in order to better resolve the peaks. Fig. 8(a) corresponds to the experimental geometry described above (*H* applied at +30°) and Fig. 8(c) shows the applied field aligned with the substrate plane (*H* applied at 0°). There is a clear difference in the demagnetizing field distribution for the two cases, and for the 0° configuration, we can see that the resulting demagnetizing field distribution is symmetric. The frequency spectrum changes drastically for this case, and it is seen to contain only two oscillation modes, which correspond to the edge and center regions. Therefore, when $H_d$ is symmetric, the spin wave modes of the cap resemble those of the flat disk in number. However, the frequencies are not the same (indicated by blue arrows) since the demagnetizing field is still different due to the curvature of the sphere. To further investigate the sensitivity of the spherical cap to the applied field direction, Fig. 8(b) shows the frequency spectrum when the field is applied at 5°. The demagnetizing field distribution again shows asymmetry, and the resulting frequency spectrum displays more than two oscillation modes. Therefore, the three-dimensional curvature of the spherical cap results in a much larger sensitivity of the magnetic oscillation modes to the applied field direction, compared to the disks, which showed two oscillation modes for all simulated field angles.

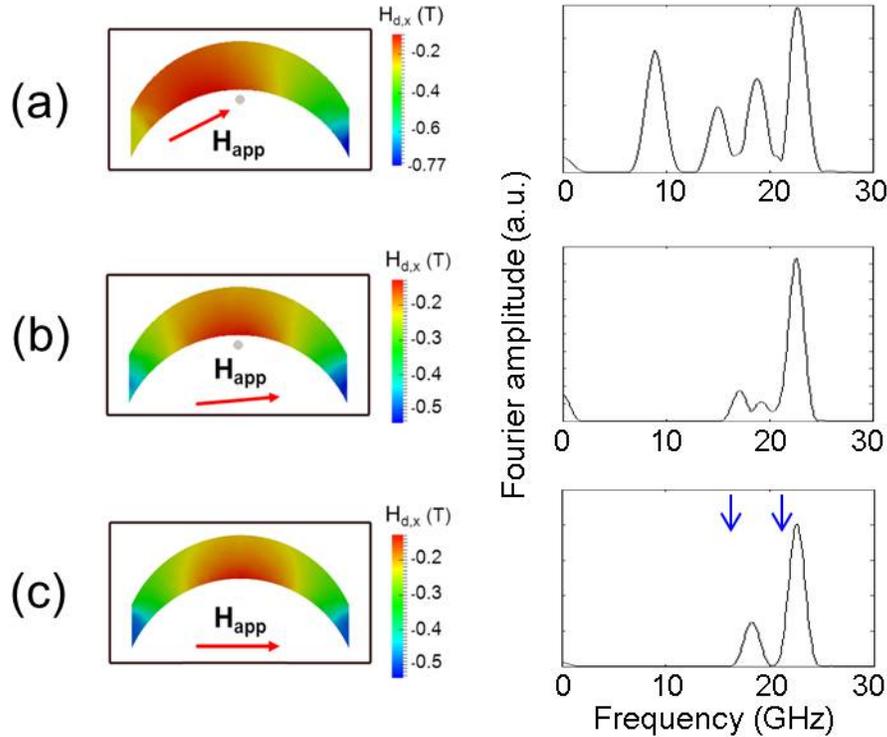

Fig. 8. (Color online) Simulated demagnetizing field distributions (left) and corresponding oscillation modes (right) for a 100 nm diameter spherical cap with different applied field angles: (a) 30° (experimental conditions), (b) 5°, and (c) 0° from the substrate plane. The applied field strength is $\mu_0 H = 0.54$ T. The blue arrows in (c) correspond to the frequencies of the flat disk spin wave modes for the same field configuration. Please note that the Fourier amplitude of the x-component of the magnetization is plotted here.

## V. Summary

We investigated the size dependence of a wide range of magnetic properties in FePt nanomagnets. We find that the three-dimensional shape is the main determining factor for quasi-static switching, as well as for the size and applied field angle dependence of optically induced picosecond dynamics.

The quasi-static switching behavior was evaluated utilizing the FORC technique. Results show that both the spherical caps and flat disks reverse by a vortex process, with the caps having smaller nucleation/annihilation fields. The annihilation field is affected by the curvature more than the nucleation field, which we attribute to a change in the demagnetization energy. Furthermore, the reduced stability of the vortex state translates into a larger critical radius for the transition from vortex annihilation to coherent reversal switching. We attribute this behavior directly to the 3D structure causing a graded radial thickness (resulting in a smaller average thickness) which changes the demagnetizing energy.

The breakup of the demagnetization field into multiple regions that results in a complex mode spectrum remains the determining factor for the magnetization dynamics over a wide range of sizes and applied field angles. As the diameter of the curved caps increases, additional pinned modes appear due to the increased area of the element which allows for the formation of additional spin wave well segments. The higher sensitivity of the spin wave modes to the applied field direction makes curved magnets attractive for flexible design the magnetization dynamics on the nanoscale.

**Acknowledgments**

This work was supported by the National Science Foundation under grants ECCS-0801896, DMR-0806924, ECCS-0925626, and DMR-1008791, the Deutsche Forschungsgemeinschaft under grant Al 618/5-1, the W. M. Keck Center for Nanoscale Optofluidics, the German Research Foundation under grants Al 618/5-1 and Al 618/11-1, and the International Research Training Group GRK 1215. The authors would like to thank M. Daniel (Chemnitz University of Technology) for RBS measurements and analysis and Y. Yahagi (UC Santa Cruz) for help with preparing the manuscript.

**References**

[1] J. Åkerman, Science **308**, 508 (2005).

[2] D. C. Worledge, G. Hu, D. W. Abraham, J. Z. Sun, P. L. Trouilloud, J. Nowak, S. Brown, M. C. Gaidis, E. J. O'Sullivan, and R. P. Robertazzi, Appl. Phys. Lett. **98**, 022501 (2011).

[3] S.H. Kang, Jour. Min. Met. Mat. Society **60**, 28 (2008).

[4] T. Thomson, G. Hu and B.D. Terris, Phys. Rev. Lett. **96**, 257204 (2006).

[5] B.D. Terris, T. Thomson, and G. Hu, Microsyst. Techn. **13**, 189 (2007).

[6] T. Thomson and B.D. Terris, in *Data Storage: Materials Perspective*, edited by S. N. Piramanayagam and T. C. Chong (John Wiley & Sons, Inc., Hoboken, NJ, 2011), pp. 256-276.

[7] I.W. Hamley, Nanotech. **14**, R39 (2003).

[8] I.L Guhr, O. Hellwig, C. Brombacher, and M. Albrecht, Phys. Rev. B **76**, 064434 (2007).

[9] M.M. Soares, E. de Baisi, L.N. Coelho, M.C. dos Santos, F.S. de Menezes, M. Knobel, L.C. Sampaio, and F. Garcia, Phys. Rev. B **77**, 224405 (2008).

[10] Y.J. Zhang, L.X. Sun, Y.X. Wang, X. Ding, Y. Cheng, and J.H. Yang, Solid State Comm. **147**, 262 (2008).

[11] J.H. Yang, N.N. Yang, Y.X. Wang, Y.J. Zhang, Y.M. Zhang, Y. Liu, M.B. Wei, Y.T. Yang, R. Wang, and S.Y. Yang, Solid State Comm. **151**, 1428 (2011).

[12] Y.J. Zhang, W. Li. J. Li, Y.M. Zhang, Y.X. Wang, S.Y. Yang, S.S. Liu, L.C. Wu, G.S.D. Beach, and J.H. Yang, J. Appl. Phys. **111**, 053925 (2012).

[13] R. Streubel, D. Makarov, F. Kronast, V. Kravchuk, M. Albrecht, and O.G. Schmidt, Phys. Rev. B **85**, 174429 (2012).

[14] M. Albrecht, G. Hu, I.L. Guhr, T.C. Ulbrich, J. Boneberg, P. Leiderer, and G. Schatz, Nat. Mat. **4**, 203 (2005).

[15] D. Makarov, E. Bermudez-Urena, O.G. Schmidt, F. Liscio, M. Maret, C. Brombacher, S. Schulze, M. Hietschold, and M. Albrecht, Appl. Phys. Lett. **93**, 153112 (2008).

[16] R. Brandt, F. Ganss, R. Rückriem, T. Senn, C. Brombacher, P. Krone, M. Albrecht, and H. Schmidt, Phys. Rev. B **86**, 094426 (2012).

[17] R. Micheletto, H. Fukuda, and M. Ohtsu, Langmuir **11**, 3333 (1995).

[18] T. Klemmer, D. Hoydick, H. Okumura, B. Zhang, and W. Soffa, ScriptaMetallurgica et Materialia **33**, 1793 (1995).

[19] N. Qureshi, S. Wang, M.A. Lowther, A.R. Hawkins, S. Kwon, A. Liddle, J. Bokor, and H. Schmidt, Nano Lett. **5**, 1413 (2005).

[20] R.K. Dumas, C.-P. Li, I.V. Roshchin, I.K. Schuller, and K. Liu, Phys. Rev. B **75**, 134405 (2007).

[21] C.R. Pike, A.P. Roberts, and K.L Verosub, J. Appl. Phys. **85**, 6660 (1999).

[22]C.R. Pike, C.A. Ross, R.T. Scaletter, and G. Zimanyi, Phys. Rev. B **71**, 134407 (2005).

[23]J. E. Davies, O. Hellwig, E. E. Fullerton, G. Denbeaux, J. B. Kortright, and K. Liu, Phys. Rev. B **70**, 224434 (2004).

[24]R. P. Cowburn, D. K. Koltsov, A. O. Adeyeye, M. E. Welland, and D. M. Tricker, Phys. Rev. Lett. **83**, 1042 (1999).

[25]R.K. Dumas, C.-P. Li, I.V. Roshchin, I.K. Schuller, and K. Liu, Phys. Rev. B **86**, 144410 (2012).

[26]J. E. Davies, J. Wu, C. Leighton, and K. Liu, Phys. Rev. B **72**, 134419 (2005).

[27]R.K. Dumas, K. Liu, C.-P. Li, I.V. Roshchin, and I.K. Schuller, Appl. Phys. Lett. **91**, 202501 (2007).

[28]W. Scholz, J. Fidler, T. Schrefl, D. Suess, R. Dittrich, H. Forster, V. Tsiantos, Comp. Mat. Sci. **28,** 366 (2003).

[29]A. Aharoni, *Introduction to the Theory of Ferromagnetism*, Oxford University Press: Oxford, 1996.

[30]H. Hoffmann and F. Steinbauer, J. Appl. Phys. **92**, 5463 (2002).

[31]J.G. Deak, IEEE Trans. Magn. **39**, 2510 (2003).

[32]B. Koopmans, in *Spin Dynamics in Confined Magnetic Structures II*, edited by B. Hillebrands and K. Ounadjela, Topics Appl. Phys. Vol. 87, (Springer-Verlag, Berlin, Heidelbert, 2003), pp. 253-320.

[33]A. Barman, S. Wang, O. Hellwig, A. Berger, E. E. Fullerton, and H. Schmidt, J. Appl. Phys. **101**, 09D102 (2007).

[34]B. Koopmans, M. van Kampen, J.T. Kohlhepp, and W.J.M de Jonge, Phys. Rev. Lett. **85**, 844 (2000).

[35]A. Barman, S. Wang, J.D. Maas, A.R. Hawkins, S. Kwon, A. Liddle, J. Bokor, and H. Schmidt, Nano. Lett. **6**, 2939 (2006).

[36]V.V. Kruglyak, A. Barman, R.J. Hicken, J.R. Childress, and J.A. Katine, Phys. Rev. B **71**, 220409R (2005).

[37]T. Schrefl, M.E. Schabes, D. Suess, O. Ertl, M. Kirschner, F. Dorfbauer, G. Hrkac, and J. Fidler, IEEE Trans. Magn. **41**, 3064 (2005).

[38]R. Rückriem, P. Krone, T. Schrefl, and M. Albrecht, Appl. Phys. Lett. **100**, 242402 (2012).

[39]P.S. Keatley, V. V. Kruglyak, A. Neudert, E.A. Galaktionov, R. J. Hicken, J.R. Childress, and J.A. Katine, Phys. Rev. B **78**, 214412 (2008).

[40]H.T. Nembach, J.M. Shaw, T.J. Silva, W.L. Johnson, S.A. Kim, R.D. McMichael, and P. Kabos, Phys. Rev. B **83**, 094427 (2011).